\documentstyle[12pt,aasms4]{article}
\begin{document}

\title{Measuring Galactic Extinction: A Test}
\author{H\'ector G. Arce\footnote{National Science Foundation
Minority Graduate Fellow} \& Alyssa A. Goodman\footnote{National Science
Foundation Young Investigator} }
\affil{Harvard--Smithsonian Center for Astrophysics, 60 Garden St.,
Cambridge, MA 02138}
\authoremail{harce@cfa.harvard.edu, agoodman@cfa.harvard.edu}
\vskip0.5in
\centerline {Accepted by {\it The Astrophysical Journal (Letters)}}

\begin{abstract}

We test the recently published
all-sky reddening map of Schlegel, Finkbeiner \&
Davis (1998 [SFD]) using the extinction study of a region in the Taurus
dark cloud complex by Arce \& Goodman (1999 [AG]).
In their study, AG use four different
techniques to measure the amount and structure of the extinction toward
Taurus, and all four
techniques agree very well. Thus
we believe that the AG
results are a truthful representation of the extinction
in the region and can be used to test the reliability of the SFD
reddening map. The results of our test
show that the SFD all-sky reddening map, which is based on data from
COBE/DIRBE and IRAS/ISSA, overestimates the reddening by a factor
of 1.3 to 1.5 in regions of smooth extinction with $A_V > 0.5$ mag.
In some regions of steep extinction gradients
the SFD map underestimates the reddening value,
probably due to its low spatial resolution.
We expect that the astronomical community will be using
the SFD reddening map extensively.
We offer this {\it Letter} as a cautionary note about
using the SFD map in regions of high extinction ($A_V > 0.5$ mag), as it might
not be giving accurate reddening values there. 


\end{abstract}

\section{Introduction}

Extinction and reddening
caused by interstellar dust effects the detected emission
from most astronomical sources in the sky.
In most galactic and extragalactic studies, especially when
studying so called ``standard candles,''
the effects of dust on the source's detected brightness and color
need to be taken into account.
Hence it is very desirable to know the
extinction (and reddening) anywhere on the sky.

Recently Schlegel, Finkbeiner and Davis (1998; hereafter SFD)
published an all-sky
reddening map, based on satellite observations of
far-infrared emission (at 100
and 240 \micron) from dust.
This new reddening map will surely be used by many researchers
seeking to apply reddening ``corrections'' for their work, so
it is very important to verify its accuracy and reliability.
The SFD map uses data obtained by
the DIRBE (Diffuse InfraRed Background Experiment) on board COBE combined
with ISSA (IRAS Sky Survey Atlas) images.  The COBE/DIRBE experiment had
better control
of absolute calibration than did IRAS, but a larger beam (0.7\arcdeg; as
compared with
$\sim$ 5\arcmin\ for IRAS). SFD use the DIRBE data to calibrate
the IRAS/ISSA images, and, after sophisticated processing, they obtain
a full sky map
of the 100 \micron \/ emission from dust, with point sources and the zodiacal
light foreground removed.\footnote {See  SFD for a full description of
their elaborate
foreground zodiacal light subtraction technique, point source extraction and
overall data reduction.}
Their resulting reddening map is more accurate and has better resolution
($\sim$ 6.1\arcmin) than the previous existing all sky reddening map of
Burstein and Heiles (1978, 1982).

In order to
test the extinction map derived from the SFD analysis, we compare it with
Arce \& Goodman's (1999; hereafter AG) recent extinction study of
a region of the Taurus dark cloud complex.
The AG study
uses four different techniques to measure extinction
along two constant-Right Ascension ``cuts'' of several degrees in length.
The four techniques in AG utilize:
1) the color excess of background stars for which spectral types are
known;
2) the ISSA 60 and 100 \micron \/ images;
3) star counts; and 4) an optical ($V$ and $R$) version
of the average color excess method
used by Lada et al. (1994).
The study finds that all four methods give generally similar results, and
concludes that
all four techniques are reliable ways to measure extinction in
regions where $A_V \lesssim 4$ mag.
In this {\it Letter} we compare the extinction map derived from SFD
to the AG extinction map which is based on ISSA 60 and 100 \micron \/ images
in the Taurus region. This comparison provides a test of the reliability of the
SFD reddening map for regions of extinction ($A_V$) higher than 0.5 mag.

\section{Comparing the two results}

In this section we compare the extinction derived from the SFD reddening map
(hereafter $A_{V_{SFD}}$) to the extinction AG obtained using the ISSA
60 and 100 \micron \/ maps 
(hereafter $A_{V_{ISSA}}$). In order to be consistent,
we must compare both
methods with the same spatial resolution.
As part of their image reduction procedure,
SFD convolve the ISSA images with a FWHM=3.2\arcmin
\/ Gaussian, which results in ISSA images with 6.1\arcmin \/ resolution.
For the purpose of comparing the two extinction maps, we also
convolve the AG ISSA images with a FWHM=3.2\arcmin \/ Gaussian, so that both
extinction determinations have the same resolution.

When AG compare $A_{V_{ISSA}}$ with the extinction measured with
star counting techniques
and the average color excess method (techniques 3 and 4 above),
they do so by plotting extinction versus declination. The
extinction values shown in Figures 3 and 5 of AG
are actually traces of the extinction averaged over the $\sim$10\arcmin \/
span (in the R.A. direction) of their CCD fields.
For consistency we consider the same kind of average, and
we obtain the SFD reddening ($E(\bv)_{SFD}$)
for the two 10\arcmin \/ wide constant--R.A.
cuts of AG and average  $E(\bv)_{SFD}$ over Right Ascension.
In order to convert from the
color excess value given by the SFD maps to visual extinction, we use
the equation $A_V=R_V E(\bv)$, where $R_V$, the ratio of total-to-selective
extinction, is equal to 3.1 (SFD;
Kenyon et al. 1994; Vrba \& Rydgren 1985).

In Figure 1 we plot extinction 
versus declination for
the two constant R.A. cuts in the Taurus dark cloud complex region.
In addition to $A_{V_{SFD}}$ and $A_{V_{ISSA}}$,
we show the extinction traces of the extinction obtained from the average color
excess and star count methods of AG, for reference.
We will concetrate our discussion on the ISSA (AG) and SFD methods. 
It can be seen that these two methods trace similar
structure in both cuts. The extinction peaks associated with the IRAS
core Tau M1 (Wood et al. 1994), the dark cloud L1506, the dark cloud
B216-217, and the IRAS cores Tau B5 and Tau B11 (Wood et al. 1994)
are easily detected.
The 1 $\sigma$ error (not including
systematics) of $A_{V_{ISSA}}$ is 0.12 mag (AG). 
SFD quote an uncertainty
of 16\% in their reddening estimates.
For the most part, in regions unassociated with a pronounced peak in
extinction,
$A_{V_{SFD}}$ is about a factor of 1.3 to 1.5 larger than $A_{V_{ISSA}}$.
It is also important to note that in
regions where there is a peak in the extinction, the value of
$A_{V_{ISSA}}$ is closer to that of $A_{V_{SFD}}$ than in the rest
of the trace, and in some of the peaks
$A_{V_{ISSA}}$ exceeds $A_{V_{SFD}}$.

There are two places where the two traces appear to show very different
extinction structures. In both of these places $A_{V_{ISSA}}$ shows
a steep dip: one is near the
peak associated with B216-217 in cut 2 (around declination 26.9\arcdeg);
and the other is near the end of the cut,
around declination 28.3\arcdeg \/ (see Figure 1).
These two regions coincide with the position of IRAS point sources,
and will be discussed in more detail further on.

\section{Comparison between the two methods}

SFD use data from the IRAS and DIRBE experiments to
construct a full-sky map of the Galactic dust based
on its far-infrared emission. The IRAS data are used as a source of
100 \micron \/ flux images, and the DIRBE data are used for absolute
calibration and as a source of
240 \micron\ data. SFD derive the dust color temperature using the ratio of
100 to
240 \micron \/ emission from DIRBE\footnote{Note that deriving dust
temperature from a long wavelength
color ratio, such as 100/240 \micron\ is superior to using the 60/100
\micron\ ratio, because the
effects of both point sources and transient heating of small grains are
minimized.}, and then use this dust color temperature to
convert their ISSA-based maps of 100 \micron \/ emission to maps of dust
column density.
As a result of this procedure, the SFD ISSA-based 100 \micron \/ emission
map has a spatial resolution
of 6.1\arcmin, but their dust color temperature map has a resolution
of $\sim$ 1.4\arcdeg \/ (Schlegel 1998).

To transform from the map proportional
to dust column (hereafter $\tau_{SFD}$)
to a reddening ($E(\bv)$) map, SFD use the correlation between
the intrinsic \bv \/ color of elliptical galaxies and the Mg$_2$
line strength.
The Mg$_2$ line strength of an elliptical galaxy
correlates well with its
{\it intrinsic} \bv \/ color so that the Mg$_2$ line index can be used along
with photometric measurements of the galaxy in order to obtain
a reasonably accurate measurement of its reddening (Faber et al. 1989).
SFD use a procedure where the residual of the
\bv \/ color versus the Mg$_2$ line index for 389 elliptical galaxies is
correlated with the estimated reddening from their maps,
using a Spearman Rank method.
The resultant fit is then used by SFD to convert from $\tau_{SFD}$
to reddening ($E(B-V)_{SFD}$) at each pixel. We then convert reddening to
extinction, using the relation $A_{V_{SFD}}=R_V E(B-V)_{SFD}$, with $R_V=3.1$
(SFD).

In order to obtain their ISSA-based extinction, $A_{V_{ISSA}}$, AG begin by using ISSA 60
and 100 \micron \/ images to obtain a dust color temperature map from the flux ratio at each
pixel.  This color temperature is then used to convert the observed 100 \micron\
flux to 100 \micron\ optical depth, $\tau_{100}$.  As in SFD, the next conversion, from dust
opacity to extinction, is tied to a separate technique of obtaining extinction.  AG chose
to use a method similar to that described in Wood et al. (1994), which itself is ultimately
based on work by Jarrett et al. (1989).  Jarrett et al. (1989) correlate 60
\micron\ optical depth ($\tau_{60}$)  with extinction ($A_V$) obtained from star counts,
and Wood et al. (1994) multiply the Jarrett et al. $\tau_{60}$ values by 100/60 to derive a
conversion from
$\tau_{100}$ to $A_V$, assuming low optical depth.  Thus, using Wood et al.'s
technique, AG's conversion of $\tau_{100}$ to $A_V$ ultimately rests on
Jarrett et al.'s correlation of 60 \micron\ optical depth with star counts.  As described in
AG, this $\tau_{60}-A_V$ correlation is very tight for $A_V<4$ mag, so we ascribe very
little of the uncertainty in $A_{V_{ISSA}}$ to this conversion.

\section{The cause of the discrepancy}

Arce \& Goodman (1999) obtain the extinction toward the Taurus region
using four different techniques.  All four give similar results in
terms of the absolute value and overall structure of the extinction. 
 Most importantly, AG
find that $A_{V_{ISSA}}$ agrees well with extinction values obtained by
measuring the color excess of background stars for which spectral types are known, which is
the the most direct and accurate way to measure reddening (see Figure 2 in AG).  So, the
question now is why does the SFD method give extinctions which differ from the four
determinations of extinction made by AG?

{\it Absolute values.} Although
data reduction and zero point determination of the emission is very
elaborate and accurate in the SFD method,
the normalization of the reddening per unit flux
density (conversion between $\tau_{SFD}$ and $E(\bv)_{SFD}$)
seems to overestimate the reddening in regions of high dust opacity.
As explained above, SFD use the correlation between intrinsic
\bv \/ color and the Mg$_2$ line index in elliptical galaxies to
convert from dust column density to reddening.
It can be seen in Figure 6 of SFD that 90\% of the 389 elliptical galaxies
they use for the \bv \/ vs. Mg$_2$ regression have $E(\bv)_{SFD}$ values
of less then 0.1 mag. Moreover it is clear
from their Figure 6 that the fit is excellent for values of  $E(\bv)_{SFD}$
less than 0.15 mag, but for $E(\bv)_{SFD} > 0.15$ mag
 the fit starts to diverge
for the few points that exist.
Galaxies with $E(\bv)_{SFD} > 0.15$ mag ($A_V > 0.5$ mag)
seem to follow a trend leading to
reddening values in regions with high dust opacity being overestimated.
SFD state that the slight trend in the residual is not
statistically significant, but that may be due to the fact that there are so
few points with $E(\bv)_{SFD} > 0.15$ mag.

Figure 6 in SFD shows only two points (galaxies) that have $E(\bv)_{SFD} > 0.3$
mag. We use the point with $E(\bv)_{SFD} = 0.32$ mag to asses the amount
of overestimation by the SFD method in high reddening regions. Using
Figure 6 in SFD and $A_V = R_V E(\bv)$, with $R_V = 3.1$, 
for an extinction of around 1 mag, the SFD method would overestimate
the extinction, at 1.31 magnitudes.
In fact, by comparing  $A_{V_{SFD}}$ to $A_{V_{ISSA}}$,
we find the overestimation by SFD can be even  more
than that: when $A_{V_{ISSA}}$ is 1,  $A_{V_{SFD}}$ is typically 1.5.
{\it Thus we are convinced
that SFD overestimates the reddening value to lines of sights where the
extinction is more than 0.5 mag}. Such overestimation is due
to the fact that in the sample of 389 galaxies
most of them have $E(\bv) < 0.1$
and very few have $E(\bv) > 0.15$.
This results in an accurate conversion between 100
\micron \/ emission and reddening for regions with very low extinction
($E(\bv) < 0.1$), but a less accurate conversion where $E(\bv) > 0.15$
($A_V > 0.5$ mag).

One could argue that the SFD method overestimates the extinction because
we overestimate the value of $R_V$, needed to convert $E(B-V)$ to $A_V$.
We believe that is not so, since several independent studies indicate that the value
of $R_V$ in the Taurus dark cloud complex is around 3.1 and that it stays
constant through the region for lines of sight in which
$A_V < 3$ mag (Kenyon et al. 1994; Vrba \& Rydgren 1985).
Furthermore, $R_V=3.1$ gives very good agreement between 
$A_{V_{ISSA}}$  and extinction obtained using the color excess of
background stars for which spectral types are known (AG).

{\it Structure.} In Section 2, we noted that there are certain places where the  traces of $A_{V_{ISSA}}$
and  $A_{V_{SFD}}$ differ not only in value but also in structure. 

The most dramatic ``structural" differences correspond to places where the
cuts in Figure 1 intercept IRAS point sources. SFD remove point sources from the
ISSA images before they use them, while AG do not.
Both AG and SFD assume a single dust temperature for each line of sight. In AG's
determination of $A_{V_{ISSA}}$, the high 60 \micron\ flux from IRAS point sources
transforms into dust with high color temperature, which in turn causes an artificial dip in
the extinction (see Wood et al. 1994 and references therein for more on this effect). We
can see that SFD have done a good job at removing point sources by noticing that where
$A_{V_{ISSA}}$ has artificial dips, $A_{V_{SFD}}$ traces the real structure (also traced by
the other methods in AG). Thus $A_{V_{SFD}}$, unlike $A_{V_{ISSA}}$,
is free of any unphysical extinction due to IRAS point sources.

Less dramatic, but more curious, structural discrepancies between the two methods are seen
near the extinction peaks in Figure 1. Although $A_{V_{SFD}}$ is in general larger than
$A_{V_{ISSA}}$ by a factor of 1.3 to 1.5, it is important to realize that
in some peaks $A_{V_{SFD}}$ is larger
by a factor less than 1.3 and in others it is smaller than $A_{V_{ISSA}}$.
In addition the gaussian-looking
extinction peaks traced by $A_{V_{SFD}}$ are wider
than those traced by $A_{V_{ISSA}}$.
Gaussian fits to the peaks associated with Tau M1 and L1506 for both
traces in cut 1 show that both peaks traced by $A_{V_{SFD}}$ have
a FWHM wider than those of the peaks traced by $A_{V_{ISSA}}$ by a factor
of 1.3. This suggests that in regions of steep extinction gradients
the reddening obtained from the SFD map has a lower resolution than the quoted 6.1\arcmin.

We believe that this ``smearing" in the SFD maps is 
caused by the difference in resolution 
between the ISSA {\it flux} maps (6.1\arcmin) and the COBE-limited {\it color
temperature} maps
(1.4\arcdeg) employed by SFD. In Figure 2 we plot the 100 \micron \/ emission ($I_{100}$)
versus declination for a section of cut 1,  where the rise in $I_{100}$
associated with Tau M1 and L1506 are clearly seen.
The plot shows the traces of $I_{100}$ obtained by
SFD ($I_{100_{SFD}}$) and by AG ($I_{100_{ISSA}}$) averaged over the $\sim$
10\arcmin \/ width (in R.A.) of the cut. It can clearly be seen that both
traces are essentially equal,
and that the peaks have the same width and height. Thus it is after both
the dust color temperature and $I_{100}$ have been used together to obtain the
extinction, that the SFD method loses spatial resolution.
Regions of steep
extinction gradients are likely to have temperature gradients, but if the
spatial resolution in the {\it dust
color temperature} is $\sim 1.4\arcdeg$ then it is unlikely that
these gradients
would be detected. If the temperature gradients
are not accounted for when calculating
reddening, the result is areas with lower effective resolution.
Therefore, the fact that SFD use a dust color temperature map with a spatial
resolution a factor of 14 larger than the $I_{100}$ map results in a
reddening map which does not have a constant spatial resolution of
6.1\arcmin. Unfortunately the DIRBE instrument has a limiting
resolution of $\sim$ 1\arcdeg, so it is not possible to do better with data
taken with that instrument.

\section{Conclusion}

We test the COBE/IRAS all-sky reddening map of Schlegel, Finkbeiner \&
Davis (1998) using the extinction study of Arce \& Goodman (1999).
In their study, AG use four different techniques to study the extinction
in a region of the Taurus dark cloud complex. All four techniques
give very similar results in terms of the value of $A_V$ and the structure
in the extinction. Thus the results of AG can be considered
a truthful representation
of the extinction in this region, and can be used to test the reliability
of the SFD reddening map.

We compare the extinction obtained by AG using
ISSA 60 and 100 \micron \/ images with the SFD reddening map. Our results show
that in general the SFD method gives extinction values a factor of 1.3 to 1.5
larger than the extinction obtained by AG.
{\it We conclude that SFD overestimates the reddening value to
lines of sights where $A_V > 0.5$ mag.}
We expect this overestimation is caused by
the fact that in the sample of 389 galaxies used to calculate a conversion from
dust column density to $E(\bv)$,
90\% of the galaxies have low reddening values ($E(\bv) < 0.1$)
and very few ($\sim 4$\%) have high reddening ($E(\bv) > 0.15$) values.
The lack of galaxies in high reddening regions
results in inaccuracy in the conversion between dust column
and reddening for lines of sight with
$E(\bv) > 0.15$. This bias should be studied in other regions of the sky
in order to see how general this trend is.
For now we advise that caution be taken when using the SFD
all-sky reddening map to obtain reddening (and extinction) values for regions
with $E(\bv) > 0.15$ ($A_V > 0.5$ mag),
as the value given by SFD could be larger than the real value.

The behavior of $A_{V_{SFD}}$ near steep gradients in extinction seems to
be different from the overall ``smoother'' extinction regions.
The difference between $A_{V_{SFD}}$
and $A_{V_{ISSA}}$ decreases near extinction peaks and in some peaks
$A_{V_{SFD}}$ even becomes smaller than $A_{V_{ISSA}}$. In addition 
the peaks in $A_{V_{SFD}}$ are wider than the $A_{V_{ISSA}}$ peaks.
We attribute this behavior to the fact that SFD use a dust
temperature map with a spatial resolution of a factor of 14 larger than
the 100 \micron \/ intensity map used to obtain the reddening map.  The poor
resolution in the temperature maps results in a reddening map which
shows less spatial resolution near extinction peaks.

\acknowledgements
We thank David Schlegel, Douglas Finkbeiner and Mark David
for their very useful comments on this work, and Krzysztof Stanek for
pointing out the SFD paper to us, and for sharing his work on this topic.
We also thank the referee for prompt and very helpful comments.

\newpage
\clearpage
\thispagestyle{empty}

\begin{figure}
\vspace{6.0in}
\centerline{
\includegraphics{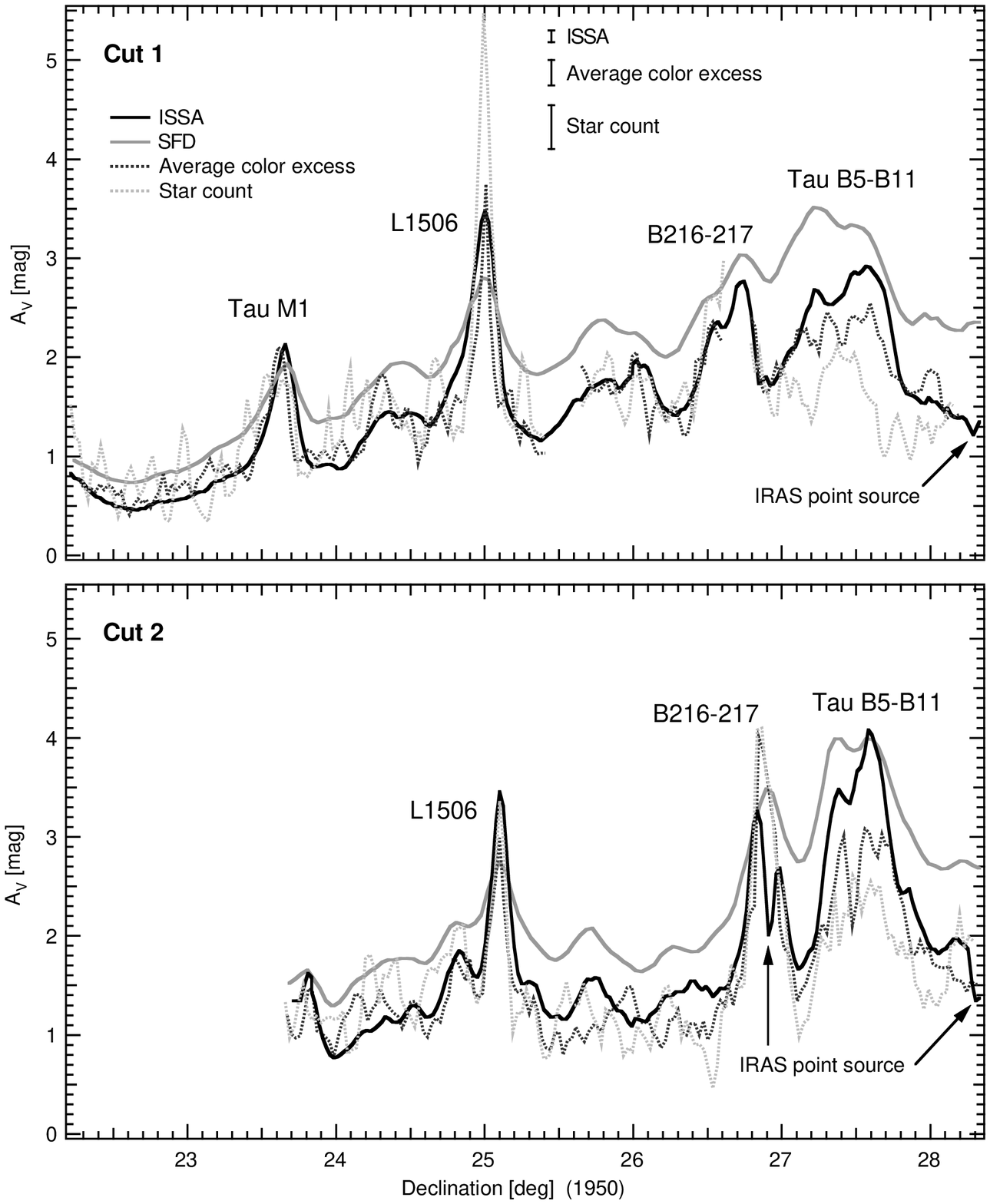}
}
\caption{Plots of extinction vs. declination for
cut 1 and cut 2 of AG. The first cut is centered on 
R.A. 4h19m33s (B1950), and the second cut
is centered on R.A.  4h18m24s (B1950).
The light solid line is the extinction obtained
from the SFD reddening map. The dark solid line
is the extinction obtained using the ISSA 60 and 100 \micron \/ images,
the light dotted line is the extinction from star counts and the 
dark dotted line is the average color excess technique from AG.
The average 1 $\sigma$ random
error for each of AG's extinction method is shown. 
The peaks in extinction  associated with named dark clouds and
IRAS cores are identified.
Dips in the $A_{V_{ISSA}}$ trace, due to artificial drops in
the extinction caused by the presence of IRAS point source in the images,
are also shown.}

\end{figure}

\newpage
\clearpage
\thispagestyle{empty}

\begin{figure}
\vspace{6.5in}
\centerline{
\includegraphics{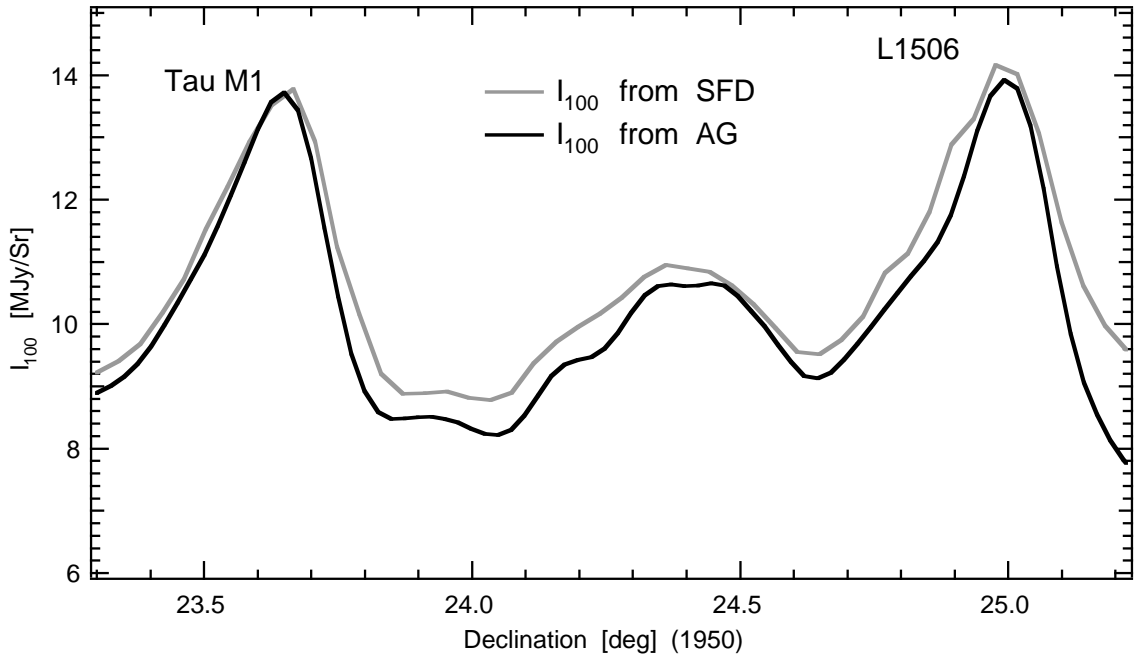}
}
\caption{A plot of 100 \micron \/ dust emission versus
declination for a part of cut 1. The light line is the emission obtained
from the SFD 100 \micron \/ map. The dark line
is the emission from the AG  100 \micron \/ map. The rise in 100 \micron \/ emission associated with Tau M1 and L1506 are identified.}

\end{figure}

\end{document}